\documentstyle[prl,twocolumn,aps,epsf]{revtex}

\begin{document}
\draft
\twocolumn[\hsize\textwidth\columnwidth\hsize\csname@twocolumnfalse%
\endcsname

\title{Critical spectral statistics in two-dimensional \\
interacting disordered systems}

\author{E. Cuevas}

\address{Departamento de F{\'\i}sica, Universidad de Murcia,
E-30071 Murcia, Spain.}

\date{\today}

\maketitle

\begin{abstract}
The effect of Coulomb and short-range interactions on the spectral
properties of two-dimensional disordered systems with two spinless
fermions is investigated by numerical scaling techniques.
The size independent universality of the critical nearest
level-spacing distribution $P(s)$ allows one to find a delocalization
transition at a critical disorder $W_{\rm c}$ for any non-zero value
of the interaction strength. At the critical point the spacings
distribution has a small-$s$ behavior $P_c(s)\propto s$, and a
Poisson-like decay at large spacings.
\end{abstract}

]
\narrowtext

It has been shown that the statistical properties of spectra of
disordered one-electron systems are closely related to the
localization properties of the corresponding wave functions
\cite{AS86,SS93,KL94}.
In the localized regime, states close in energy have an
exponentially small overlap and their levels are uncorrelated. So
the corresponding normalized spacings $s$ are distributed according
to the Poisson law
\begin{equation}
P_{\rm P}(s) = \exp(-s)\;.\label{poisson}
\end{equation}
On the other hand, in the metallic regime the large overlap of
delocalized states induces correlations in the
spectrum leading to the well known level repulsion effect. If the
system is invariant under rotations and under time-reversal symmetry
the spacings follow Wigner-Dyson statistics
\begin{equation}
P_{\rm W}(s) = \frac{\pi}{2}s\exp \left (-\frac{\pi}{4}s^2 \right )\;.
\label{wigner}
\end{equation}
The two limiting behaviors given by Eqs.\ (\ref{poisson}) and
(\ref{wigner}) correspond to an infinite system.
In Refs.\ \cite{SS93,AZ88} was found that a third, size independent
distribution, appears at the metal-insulator transition (MIT) in the
three dimensional (3D) Anderson model. This behavior is the basis of
a different method to detect the MIT whose advantage resides in that
no knowledge of the eigenfunctions is needed.
We will extend this
procedure to two-electron interacting disordered systems.

The study of the energy spectrum has already provided important results
about the nature of the states and about the transport regimes that one
can expect.
In fact, the method has been successful in the location of the MIT in
the 3D Anderson model \cite{SS93,HS94,Ho98,ZK95,ZK97},
in two dimensional (2D) disordered systems with symplectic symmetry
\cite{Ev95,SZ95,SZ97}, in the determination of the critical interaction
strength in many-body systems \cite{BA98,Ja98,JS97},
and in the study of transport regimes in the 2D Anderson model
\cite{ZB96,CO98}.

As dimension two is marginal, one can expect a interaction-driven
delocalization transition in a 2D interacting
disordered system because of the following reasons.
First of all, the scaling theory of localization including the combined
effects of disorder and interactions predicts that a 2D system may
remain metallic even in the limit of zero temperature \cite{Fi83,CC98}.
Secondly, many recent experimental results have presented strong
evidence for a delocalization transition
in 2D systems without magnetic scattering effects employing
different materials and designs \cite{KK96,KM95,KS96,PF97,PW97,Si98,Ha98}.
All these experiments show clear indications that strong
electron-electron interactions partially suppress
the quantum interference effects responsible for localization.
Furthermore, a recent numerical study of two-interacting bosons in
2D by means of their decay length in long bars \cite{OC98} showed
evidence for a localized to extended states transition.

In this Letter we study the level statistics of two interacting
electrons in a 2D random potential. We consider both long-range
Coulomb interactions and short-range interactions and calculate
numerically the correlations in the exact two-electron spectrum.
Our main result is that the nearest level-spacing distribution
$P(s)$ is universal in the critical region showing a delocalization
transition for any finite value of the interaction strength.

We consider spinless fermions on a sample of size $L \times L$
described by the standard Anderson-Hubbard Hamiltonian
\begin{eqnarray}
H & = & t\sum_{i,j} (a_{i,j+1}^{\dagger}a_{i,j}+ a_{i+1,j}^{\dagger}
a_{i,j}+ {\rm h.c.}) \nonumber \\
& & + \sum_{i,j} \epsilon_{i,j}a_{i,j}^{\dagger}a_{i,j}+H_{\rm int} \;,
\label{hamil}
\end{eqnarray}
where the operator $a_{i,j}^{\dagger}$ ($a_{i,j}$) creates (destroys)
an electron at site $(i,j)$ of a square lattice and $\epsilon_{i,j}$ is
the energy of this site chosen randomly between $(-W/2, W/2)$ with
uniform probability.
The hopping matrix element $t$ is taken equal to $-1$ and the lattice
constant equal to 1, which sets the energy and length scales,
respectively.
For the long-range case, the interaction Hamiltonian is given
by \cite{BA98}
\begin{equation}
H_{\rm int}=U\sum_{i,j>k,l}\frac {a_{i,j}^{\dagger}
a_{i,j}a_{k,l}^{\dagger} a_{k,l}} 
{|\makebox{\boldmath $r$}_{i,j} - \makebox{\boldmath $r$}_{k,l}|}\;,
\label{longri}
\end{equation}
while for the short-range case we choose a nearest-neighbor interaction
Hamiltonian
\begin{eqnarray}
H_{\rm int} &=& U\sum_{i,j} (a_{i,j}^{\dagger}a_{i,j}a_{i,j+1}^{\dagger}
a_{i,j+1} \nonumber \\
& &  +a_{i,j}^{\dagger}a_{i,j}a_{i+1,j}^{\dagger}a_{i+1,j})\;.
\label{shortri}
\end{eqnarray}
In both cases, in order to reduce edge effects, we use periodic
boundary conditions,
{\it i.e.}, $a_{L+1,j}^{\dagger}=a_{1,j}^{\dagger}$,
$a_{i,L+1}^{\dagger}=a_{i,1}^{\dagger}$,
$a_{0,j}^{\dagger}=a_{L,j}^{\dagger}$,
$a_{i,0}^{\dagger}=a_{i,L}^{\dagger}$.

The $L^4$ eigenstates of the Hamiltonian (\ref{hamil}) are either
symmetric or antisymmetric with respect to the interchange of the
electron positions. We restrict our investigation to the two-electron
Hilbert subspace spanned in the basis of $N=L^2(L^2-1)/2$ antisymmetric
products of one-electron states 
\begin{equation}
|\psi_{i,j;k,l} \rangle=\frac{1}{\sqrt {2}}(a_{i,j}^{\dagger}
a_{k,l}^{\dagger} - a_{k,l}^{\dagger}a_{i,j}^{\dagger}) |0 \rangle \;,
\end{equation}
where $|0 \rangle$ is the vacuum state.
The resulting $N \times N$ Hamiltonian matrix is numerically
diagonalized using techniques for large
sparse matrices, in particular a Lanczos tridiagonalization without
reorthogonalization method\cite{CW85}.
The strength $U$ of the interactions is varied between $0-10$
and the disorder strength ranges the interval $(0.2,16)$.
The system size varies from $L=6$ to $20$ and the number of random
realizations is such that for a given triad of $\{ U,L,W \}$ the number
of studied eigenvalues in the range $(-1,1)$ was kept around
$2.5 \times 10^4$.

In order to minimize fluctuations in $P(s)$, we used in the calculations
for $U \agt 1$ the levels contained in the interval $(-1,1)$ around the
band center.
Although this energy window exceeds the one-electron level spacing
$\Delta_1$, does not introduce hidden correlations (unlike for
the non-interacting case $U=0$ or for $U \ll 1$) in the two-electron
spectrum because of the strong mixing of the states directly coupled
by the interaction.
In Ref. \cite{WP97} was shown, modeling the full Hamiltonian
(\ref{hamil}) by a Gaussian ensemble of random matrices with a
preferential basis \cite{WP96,PS94}, that the spread width $\Gamma_2$
of a two-electron state is of the order of $\Delta_1$.
In any case, we verified that, within the statistical errors, there is
no significant changes in $P(s)$ if we used an energy window
$\Delta_{\epsilon} < \Delta_1$.

To quantitatively analyze the scaling of the statistical properties of
the eigenvalue spectrum one can use one of the two following variables
widely employed. The first one \cite{SS93,ZK95,BA98,Ja98,JS97} is given
by $\eta=\int_{s_0}^{\infty} [P(s)-P_{\rm W}(s)]ds /
\int_{s_0}^{\infty} [P_{\rm P}(s)-P_{\rm W}(s)]ds$, where 
for $s_0$ one can choose one of the two crossing points of the curves
$P_{\rm P}(s)$ and $P_{\rm W}(s)$, namely $s_0=0.473$ and $s_0=2.002$.
The former is representative of the small-$s$ part of $P(s)$, whereas
the latter is representative for its tail.
The second variable \cite{HS94,Ho98,HS93},
based in the $\Delta_3$ statistics, which take into account
correlations between non adjacent levels, is given by
$\eta=\frac{1}{30} \int_{0}^{30} \Delta_3 (L) dL$.
In this Letter we use a different scaling variable \cite{CO98,CL96}
based on the variance of $P(s)$, since it contains information about
the whole distribution, namely
\begin{equation}
\eta (L,W) = \frac{ {\rm var} (s)- 0.273 }{1-0.273 }\;,\label{eta}
\end{equation}
which describes the relative deviation of ${\rm var} (s)$
from the Wigner-Dyson limit due to the finiteness of the system.
In Eq.\ (\ref{eta}) ${\rm var} (s)=\langle s^2\rangle-\langle s
\rangle^2$, and 0.273 and 1 are the variances of Wigner-Dyson and
Poisson distributions, respectively. In this way, $\eta (L,W)$ ranges
from 0 [$P(s)=P_{\rm W}$] to 1 [$P(s)=P_{\rm P}$].
We have checked that our results are basically independent of 
which of the three definitions of $\eta$ mentioned is used, although
the results based on Eq.\ (\ref{eta}) are more accurate.

Figure\ 1 shows the disorder dependence of $\eta$ for a long-range
interaction with $U=1$ for different system sizes: $L=6$ ($\circ$),
9 ($\Box$), 16 ($\diamond$) and 20 ($\bigtriangledown$).
One can distinguish three different regions, in which $\eta$ increases,
remain constant, or decreases with increasing $L$.
The curves for different sizes cross at a common point, $W_c$, which
corresponds to the interaction-induced delocalization transition.

\begin{figure}
\begin{picture}(220,210) (-55,-50)
\epsfbox{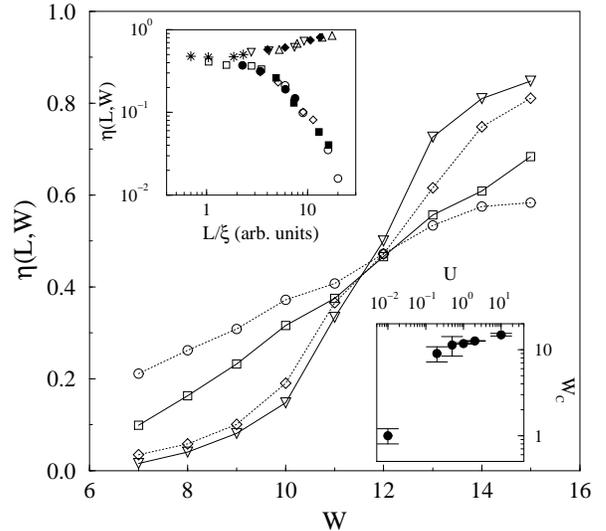}
\end{picture}
\caption{Scaling variable $\eta$ as a function of $W$ for a long-range
interaction with $U=1$. System sizes: $L=6$ ($\circ$), 9 ($\Box$), 16
($\diamond$) and 20 ($\bigtriangledown$).
The critical disorder is given by the point for which $\eta$
is independent of $L$.
Upper inset: Verification of the one-parameter scaling assumption for
$\eta(L,W)$. The horizontal scale is arbitrary.
Lower inset: $U$ dependence of $W_c$.
\label{fig.1}}
\end{figure}

In the following we check whether $\eta(L,W)$ can be expressed by
a one-parameter scaling function
\begin{equation}
\eta(L,W)=f(L/\xi (W))\;,\label{scaling}
\end{equation}
where the scaling parameter $\xi$ is the two-electron localization length
in the localized regime, and the two-particle correlation length in the
extended regime.
To perform this scaling procedure with the $\eta$ data, the range of
$\eta$ values for different $L$ at any given disorder $W_1$ must overlap
the range of $\eta$ values for various $L$ for at least one different
disorder $W_2$.
Equation (\ref{scaling}) implies that in a log-log plot of $\eta(L,W)$
versus $L$ all data should collapse in a common curve when translated by
an amount $\ln \xi(W)$ along the horizontal axis.
This curve has a single branch when there is no transition, while it
develops two separate branches when a transition is present.
In the upper inset of Fig.\ 1, we re-plot the data of Fig.\ 1 by rescaling
$L$. Note that one can fit most of the points onto a common curve with two
branches, one growing ($W > W_c$) and another decaying ($W < W_c$).
This results confirm the scaling hypothesis for $\eta(L,W)$
and thus we have found the function $\xi (W)$ up to an arbitrary factor.

Next we determine the critical disorder $W_c$ and the critical exponent
$\nu$. For this purpose we used the singularity of the correlation length
near $W_c$
\begin{equation}
\xi(W) = \xi_0 |W-W_c|^{-\nu}\;, \label{xiw}
\end{equation}
where $\xi_0$ is a constant.
Using Eq.\ (\ref{scaling}) and taking into account that
$\eta(L,W)$ is analytical for a finite system, we can write
around the critical point
\begin{equation}
\eta(L,W) = \eta_c+ \sum_{n}A_n(W-W_c)^nL^{n/\nu}\;, \label{fit}
\label{alfa1}
\end{equation}
from which one can extract $W_c$ and $\nu$. 
In practice, we have truncated the series at $n=4$.
We performed an statistical analysis of the data in the range
$10 \le W \le 14$ with the Levenberg-Marquardt method for nonlinear
least-squares models.
The most likely fit is determined by minimizing the $\chi^2$ statistic
of the fitting function (\ref{fit}).
The critical disorder found is $W_{\rm c}=11.8\pm 0.2$, and the
corresponding critical exponent is equal to $\nu=1.2\pm 0.2$.

We repeated the same kind of calculations for different values
of the interaction strength $U$ including the non-interacting case
($U=0$). For $U \ll 1$ we obtained $\eta$ using Eq.\ (\ref{eta})
in a energy window $\Delta_{\epsilon} < \Delta_1$.
The $U$ dependence of the critical disorder $W_c$ is shown in the
lower inset of Fig.\ 1. The $W_c$ found increases monotonously with
$U$ starting from very small values. In the non-interacting limit
there is no critical behavior and all states are localized,
as expected.

We have also analyzed the behavior of two electrons subjected to
a short-range interaction, Eq.\ (\ref{shortri}). The corresponding
results are depicted if Fig.\ 2. The system sizes used are the same
as those appearing in Fig.\ 1. For $U=1$ the critical disorder
and exponent found are $W_c=10.4 \pm 0.2$  and $1.3 \pm 0.2$,
respectively.
The dependence of $W_c$ on $U$, shown in the lower inset of Fig.\ 2,
is slightly different than that found for the long-range case. $W_c$
increases with $U$ and reaches a saturation value
around $U \approx 2$.
After saturation $W_c$ decreases for the largest value of the
interaction strength used, $U=10$. The reason is that for this
value of $U$, as we have checked, the interaction slightly splits
the two-interacting particle band  into lower and upper
Hubbard bands.

\begin{figure}
\begin{picture}(220,210) (-55,-50)
\epsfbox{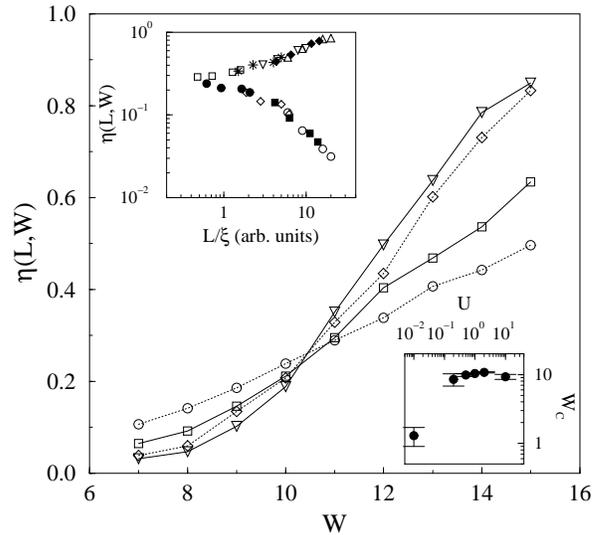}
\end{picture}
\caption{Same as in Fig.\ 1 for a short-range interaction with $U=1$.
\label{fig.2}}
\end{figure}

Finally we focus on the level statistics at the critical point.
For the orthogonal symmetry $P_c (s) \propto s$ as $s\to 0$
\cite{SS93,Me91}, while for $s \gg 1$, two different analytical
expression were proposed.
One of them \cite{SS93,AZ88} assumes that $P_c (s)$ is Poissonian and
the other \cite{AK94} has the asymptotic form
$P_c (s) \propto \exp (-A_c s^{\alpha})$, where $\alpha$ is related to
the critical exponent $\nu$ and to the dimensionality $d$ through
$\alpha = 1+(d \nu)^{-1}$.
In order to diminish the magnitude of the relative fluctuations and to
analyze the asymptotic behavior in detail, it is more convenient to
consider the cumulative level spacing distribution function
$I(s)=\int_{s}^{\infty} P(s')ds'$.
The Poisson distribution Eq.\ (\ref{poisson})
and the Wigner surmise Eq.\ (\ref{wigner}) yield $I_{\rm P}(s)=\exp (-s)$
and $I_{\rm W} (s)=\exp (-\pi s^2/4)$, respectively.

We have checked that for various $L$ considered at the critical disorder
$W_c$ for a given $U$, the best fit using a $\chi^2$ criterion in the
interval $2<s<5$ yields $\alpha \approx 1.0\pm 0.1$ and, therefore,
$P_c(s)$ at large $s$ is very close to a Poissonian decay, thus confirming
the ideas of \cite{SS93,AZ88}.
A similar conclusion was obtained in Ref.\ \cite{ZK97} in the Anderson
transition using very large system sizes.
The results for the numerical calculations of the critical $I_c (s)$
for a long-range interaction with $U=1$ are shown in Fig.\ 3 for $L=12$
($\circ$), 16 ($\Box$) and 20 ($\diamond$). Note that the critical
$I_c(s)$ is a $L$-independent universal scale-invariant function that
interpolates between Wigner and Poisson limits.
Solid and dashed lines correspond to $I_{\rm P}(s)$ and
$I_{\rm W}(s)$, respectively.
In the upper inset we display our results for $I_c(s)$ in the limit
$s\gg 1$. One clearly see that the Poissonian tail is recovered for
large spacings. This behavior is similar to that in the insulating regime,
although the decay rate $A_c$ is larger than unity. The straight line
fitting the data in the interval $2<s<5$ is $\ln I_c(s)=-1.32s+0.33$.
In the lower inset we show the limiting behavior of $P_c (s)$ as
$s \to 0$. Here we find that $P_c (s)\propto s$ according with
predictions of Refs.\ \cite{SS93,Me91}. The straight line fitting
the data in the interval $4\times 10^{-3}<s<10^{-1}$ is
$P_c (s)=3s$.

\begin{figure}
\begin{picture}(220,210) (-55,-50)
\epsfbox{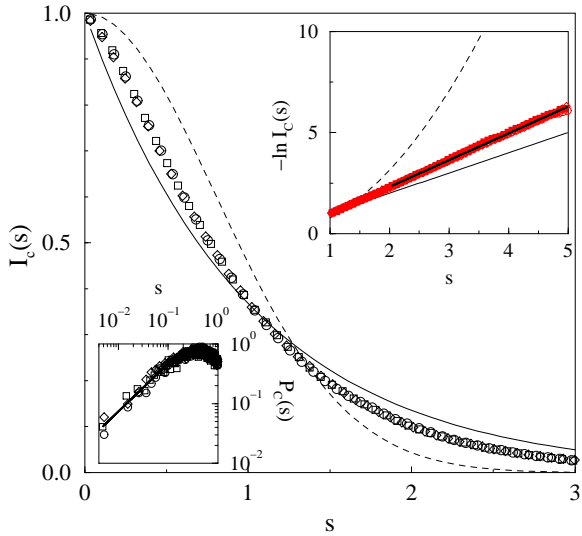}
\end{picture}
\caption{Integrated probability $I_c(s)$ at the critical disorder
for a long-range interaction with $U=1$. System size: $L=12$ ($\circ$),
16 ($\Box$), 20 ($\diamond$). Solid and dashed lines are $I_{\rm P}(s)$
and $I_{\rm W}(s)$, respectively. Upper inset: Large-$s$ part of $I_c(s)$;
the fitted line is $\ln I_c(s)=-1.32s+ 0.33$.
Lower inset: double logarithmic plot of $P_c (s)$ for $s \ll 1$;
the fitted line is $P_c(s)=3s$.
\label{fig.3}}
\end{figure}

To summarize, we have investigated the energy
levels statistics of a two-electron disordered 2D system
with a long-range or a short-range interaction. In both cases, the
scaling function was calculated using the statistics of the
nearest-neighbor eigenvalues of the two-electron states, that have
been obtained numerically.
By performing a finite size scaling analysis we found a critical
disorder $W_c$, for any non-zero value of the interaction strength $U$.
When $U \to 0$, $W_c \to 0$ and so we recover the expected results of
the non-interacting system where all the states are localized.
$P_c(s)$ is linear for small $s$ while the large-$s$ part of
$P_c(s)$ obtained is shown to have a Poisson-like decay.

We would like to thank M. Ortu\~no and D. Weinmann for
useful discussions. The Spanish DGES, project number PB96-1118, is
also acknowledge for financial support.

\end{document}